\documentclass{IEEEtran}
\usepackage{bm}
\usepackage{comment}
\usepackage{balance}
\let\temp\rmdefault
\usepackage{mathpazo}
\let\rmdefault\temp
\usepackage{amsmath}
\usepackage{amsthm}
\usepackage{amssymb}
\usepackage{mathtools}
\usepackage[table,xcdraw]{xcolor}
\usepackage{booktabs}
\usepackage{tabularx}
\usepackage[caption=false]{subfig}
\usepackage{cite}
\usepackage{url}	
\usepackage{soul}
\newtheorem{theorem}{Theorem}

\DeclareMathOperator*{\argmin}{arg\,min}
\DeclareMathOperator{\tr}{tr}

\newcommand{\RN}[1]{\textup{\uppercase\expandafter{\romannumeral#1}}}

\usepackage[acronym,style=alttree,nomain,shortcuts,nonumberlist,nogroupskip]{glossaries}
\usepackage[acronym,style=alttree,nomain,shortcuts,nonumberlist,nogroupskip]{glossaries}
\newacronym{fa}{FA}{false alarm}
\newacronym{tta}{TTA}{time to authenticate}
\newacronym{det}{DET}{detection error tradeoff}
\newacronym{md}{MD}{missed detection}
\newacronym{mse}{MSE}{mean square error}
\newacronym{cdf}{CDF}{cumulative distribution function}
\newacronym{gnss}{GNSS}{global navigation satellite system}
\newacronym{nma}{NMA}{navigation message authentication}
\newacronym{osnma}{OS-NMA}{open service navigation message authentication}
\newacronym{has}{HAS}{high accuracy service}
\newacronym{sv}{SV}{space vehicle}
\newacronym{cdma}{CDMA}{code division multiple access}
\newacronym{pvt}{PVT}{position, velocity, and time}
\newacronym{pnt}{PNT}{position navigation and timing}
\newacronym{nmea}{NMEA}{national marine electronics association}
\newacronym{rinex}{RINEX}{receiver independent exchange format}
\newacronym{sdr}{SDR}{software-defined radio}
\newacronym{cas}{CAS}{commercial authentication service}
\newacronym{acas}{ACAS}{assisted \ac{cas}}
\newacronym{bpsk}{BPSK}{binary phase-shift keying}
\newacronym{los}{LOS}{line-of-sight}
\newacronym{ls}{LS}{least squares}
\newacronym{tesla}{TESLA}{timed-efficient stream loss-tolerant authentication} 
\newacronym{sce}{SCE}{spreading code encryption}
\newacronym{sca}{SCA}{spreading code authentication}
\newacronym{prn}{PRN}{pseudo-random noise}
\newacronym{sqm}{SQM}{signal quality monitoring}
\newacronym{chimera}{CHIMERA}{chips-message robust authentication}
\newacronym{sssc}{SSSCs}{spread spectrum security codes}
\newacronym{imu}{IMU}{inertial measurement unit}
\newacronym{mac}{MAC}{message authentication code}
\newacronym{ntp}{NTP}{network time protocol}
\newacronym{ptp}{PTP}{precision time protocol}
\newacronym{scer}{SCER}{secure code estimation and replay}
\newacronym{gsc}{GSC}{GNSS service center}
\newacronym{recs}{RECS}{re-encrypted code sequence}
\newacronym{ecs}{ECS}{encrypted code sequence}
\newacronym{glrt}{GLRT}{generalized likelihood ratio test}
\newacronym{lrt}{LRT}{likelihood ratio test}
\newacronym{boc}{BOC}{binary offset carrier}
\newacronym{ppm}{ppm}{parts per million}
\newacronym{utc}{UTC}{coordinated universal time}
\newacronym{gst}{GST}{Galileo system time}
\newacronym{iot}{IoT}{Internet of things}
\newacronym{bgd}{BGD}{broadcast group delay}
\newacronym{lan}{LAN}{local area network}
\newacronym{sbf}{SBF}{Septentrio binary format}
\newacronym{vctcxo}{VCTCXO}{voltage-controlled and temperature-controlled crystal oscillator}
\newacronym{cs}{CS}{commercial service}
\newacronym{roc}{ROC}{receiver operating characteristic}
\newacronym{awgn}{AWGN}{additive white Gaussian noise}
\newacronym{kl}{K-L}{Kullback-Leibler}
\newacronym{snr}{SNR}{signal to noise ratio}
\newacronym{pdf}{pdf}{probability density function}
\newacronym{an}{AN}{artificial noise}

\begin{document}

\title{On the Optimal Spoofing Attack and Countermeasure in Satellite Navigation Systems}

\author{ \IEEEauthorblockN{Laura~Crosara, Francesco~Ardizzon, Stefano~Tomasin, Nicola~Laurenti}\\
\vspace{1mm}%
\normalsize
\IEEEauthorblockA{Department of Information Engineering, University of Padova, Italy}\\

\normalsize
 \IEEEauthorblockA{Email: \texttt{\{crosaralau, ardizzonfr, tomasin, nil\}@dei.unipd.it }}
}

\maketitle

\begin{abstract}
The threat of signal spoofing attacks against \ac{gnss} has grown in recent years and has motivated the study of anti-spoofing techniques. However, defense methods have been designed only against specific attacks. This paper introduces a general model of the spoofing attack framework in \ac{gnss}, from which optimal attack and defense strategies are derived. We consider a scenario with a legitimate receiver (Bob) testing if the received signals come from multiple legitimate space vehicles (Alice) or from an attack device (Eve). We first derive the optimal attack strategy against a Gaussian transmission from Alice, by minimizing an outer bound on the achievable error probability region of the spoofing detection test. Then, framing the spoofing and its detection as an adversarial game, we show that the Gaussian transmission and the corresponding optimal attack constitute a Nash equilibrium. Lastly, we consider the case of practical modulation schemes for Alice and derive the generalized likelihood ratio test.
Numerical results validate the analytical derivations and show that the bound on the achievable error region is representative of the actual performance.

\end{abstract}


\IEEEpeerreviewmaketitle

\glsresetall
\section{Introduction}
\begin{figure*}
\vspace{0.5cm}
\centering
\includegraphics[width=\textwidth]{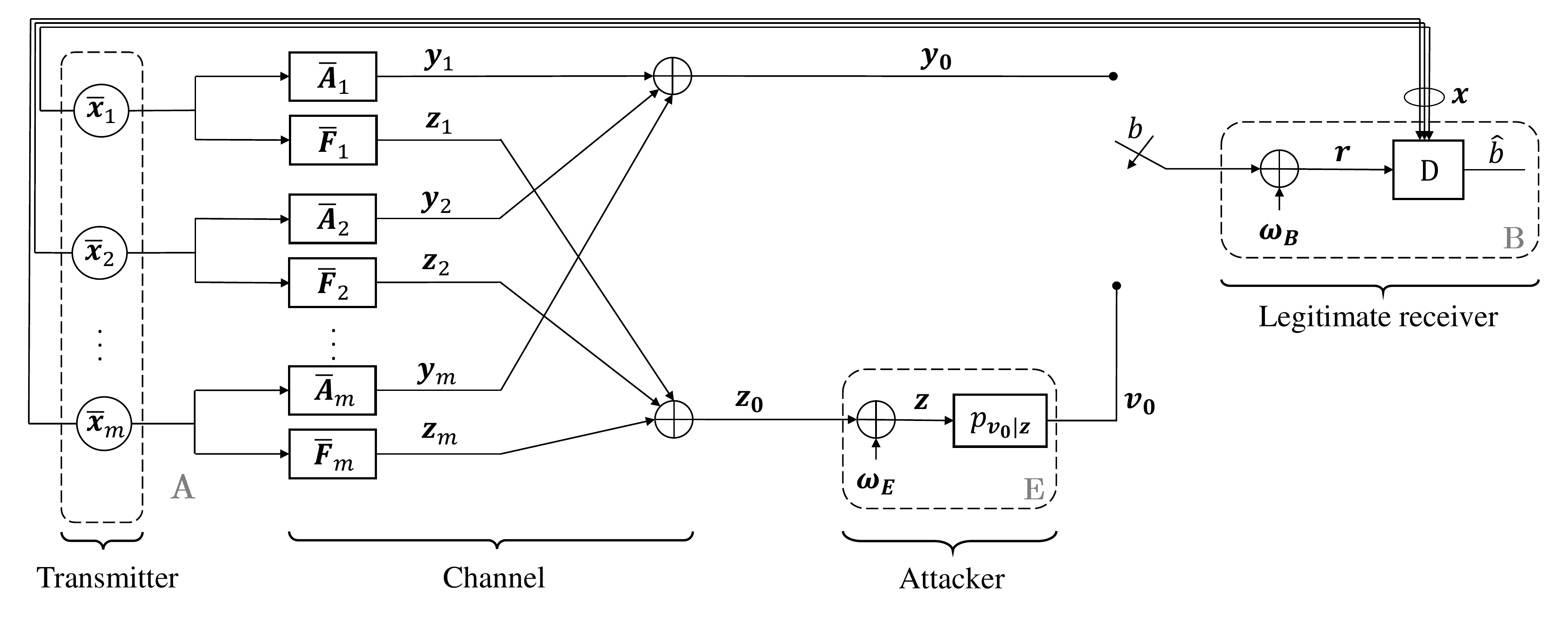}
\caption{Anti-spoofing authentication model.}
\label{fig:Model}
\end{figure*}
A growing number of location based services rely on \acp{gnss} for positioning and timing, but the widespread adoption of \acp{gnss} has also increased the incentive to mount attacks against them \cite{GNSSsurvey}. In particular, the \textit{spoofing attack} refers to the transmission of counterfeit \ac{gnss}-like signals with the intent to produce a wrong position computation at the receiver \cite{motella,GNSSsurvey,psiaki}. One of the simplest spoofing techniques is \textit{meaconing}, i.e., the re-transmission of the signal received by the spoofer towards the victim, so that the latter computes the ranging estimate based on the signals seen from the spoofer location. Advanced versions of this attack selectively forge delayed versions of the ranging signals, allowing the spoofer to induce an arbitrary position estimate at the victim. When dealing with ranging signals protected by cryptography, the \ac{scer} attack \cite{HumDetStrat,improvingSCER} can be used to estimate the message signature and then reconstruct the signal in real-time based on the estimation. Another replay attack type, called distance-decreasing attack \cite{distDecr}, modifies the receiver’s position by decreasing the pseudo-range estimation, attempting to decode a received symbol before the end of its duration.

In the last decade, the \ac{gnss} community has investigated anti-spoofing techniques both operating at data level and at ranging code level. Securing the range estimation means authenticating the source and protecting the integrity of the received signal, which requires to act at the physical layer. Therefore, in this paper, we focus on physical-layer authentication, which exploits the communication medium and does not rely on a higher layer encryption.

\Ac{sce} is the most reliable option to limit access to \ac{gnss} signals as it makes the spreading code fully unpredictable for the attacker, therefore limiting its capabilities to perform a successful signal generation attack \cite{CurranSecuringGNSS}. Moreover, when using \ac{sce}, a \ac{scer} attack has limited success in estimating each spreading code chip from the noisy received signal, since the chip period is typically several orders of magnitude lower than the message symbol period \cite{caparra18}. 
Some \ac{sce} solutions are the P(Y) code for GPS and the \ac{cas} for Galileo. \Ac{cas} is currently under development, 
but is expected to be established by 2024: in particular, a proposal known as \ac{acas}, recently presented in \cite{acasICL,cas22}, provides a \ac{sce} method where the encryption key is derived from the unpredictable signature in the Galileo \ac{osnma} \cite{OSNMAart2,OSNMAart}.

A modification of the \ac{sce} approach for civilian signals is proposed in \cite{kuhn}, where a \ac{sca} technique, which makes use of signal watermarking, is outlined. A similar \ac{sca} technique was also presented in \cite{Scott2003}, where short sequences of spread spectrum security codes (SSSCs) are used to modify the spreading code. This design approach has evolved in \cite{Scott2013}, and has been applied in \cite{chimera,chimeraION}, where the scheme called \ac{chimera} is introduced, aiming at jointly authenticating both the navigation data and the spreading code of GPS signals for civil usage.
Another solution for the joint authentication of navigation data and spreading code chips, referred to as spreading code and navigation data based authentication proposal (SNAP), can be found in \cite{snap} and a similar approach was also proposed in \cite{CurranSecuringGNSS}.
Other examples of \ac{sca} techniques for open \ac{gnss} signals, each employing different methods for generating and placing the watermarked chips, were introduced in \cite{binaryph,pozzobonSAS,pozzobonSupersonic}. 
However, existing anti-spoofing mechanisms in the literature are designed as a particular solution without any optimality criterion and their performance is evaluated against some  specific attacks, which may not represent the worst case scenario for the design mechanism.

A first unified general model for the design, description, evaluation, and comparison of \ac{sca} techniques was presented in \cite{poltronieriICL}, where an optimal compromise is achieved between security and the number of random bits in the watermark. However, the security is only evaluated in terms of conditional guessing probability for the watermarked code given the public one, thereby neglecting the effects of channel transmission, noise, and possible \ac{scer} attacks.   
On the other hand, in \cite{ANTomasin} a signal authentication method based on physical layer secrecy is presented, in which the navigation signal is transmitted along with a synchronous and orthogonal authentication signal, which is in turn encoded for secrecy and corrupted by \ac{an}. A general security result follows from the impossibility for a spoofer to decode the authentication signal without knowledge of the \ac{an}, which is later disclosed for verification. 
 
However, a totally general framework for deriving a wide class of solutions, optimizing their parameters, and evaluating their security against a broad set of attacks is still lacking, to the best of our knowledge, and is the aim of the present work.

This paper makes the following contributions. First, we describe a general model to characterize the spoofing attack in \ac{gnss}, considering the presence of multiple \acp{sv}, a victim receiver, and an attacker. 
We describe the optimal attack strategy that minimizes the \ac{kl} divergence between the received signal distribution in nominal and attack conditions, while still introducing the desired shifts on the satellites' signals. Indeed, the \ac{kl} divergence gives an outer bound to the \ac{det} curve (determined by \textit{false alarm} and \textit{missed detection} probabilities), which in turn is the appropriate metric to assess the capabilities of the victim receiver in detecting the attack.
Therefore, the proposed optimization procedure provides an optimal attack abstracting from the particular detection process and whose efficacy can be assessed beforehand.

For a general class of attack strategies, we derive a closed form expression for the \ac{kl} divergence, and a lower bound that only depends on the \ac{gnss} scenario and channel physical parameters, regardless of the specific detection strategy adopted by the victim.
Then, framing the spoofing and its detection as an adversarial game, we prove that the set of strategies comprised of a Gaussian transmission and the optimal attack is a Nash equilibrium for the game. Moreover, we discuss the \ac{kl} divergence obtained at the equilibrium points. 
Lastly, we consider the case of practical modulation schemes for Alice and derive the generalized likelihood ratio test. 
Finally, simulation results for the attack-defense scheme are presented, considering both \ac{lrt} and \ac{glrt} attack detection mechanisms, to validate the analytical derivations and show that the bound on the achievable error region given by the \ac{kl} divergence is representative of the actual performance.

The rest of this paper is organized as follows. Section~\ref{par:model} presents the general \ac{gnss} spoofing model in details, together with performance metrics. In Section~\ref{par:min} the optimal attack strategy is derived, then analytical results are presented.
Defense strategies at both the transmitter and the receiver are discussed in Section~\ref{par:defense}. 
Then, numerical results are presented in Section~\ref{par:results}, employing two transmission modulations: Gaussian signaling and finite-cardinality signaling. 
Lastly, Section~\ref{par:conclusion} draws the conclusions of the paper.

\paragraph*{Notation}
Vectors and matrices are denoted by lower and upper case boldface letters, respectively.
Symbol $\mathbold{A}^\mathrm{H}$ denotes the complex conjugate transpose of matrix $\mathbold{A}$, while $\mathbold{A}^\dagger$ denotes the Moore-Penrose pseudo inverse of $\mathbold{A}$. Symbol $\mathbold{I}_n$, denotes the identity matrix of size $n \times n$,  $|\mathbold{A}|$ and $\|\mathbold{A}\|_F$ stand for the determinant and the Frobenius norm of $\mathbold{A}$, respectively. 
Given two random variables $x$ and $y$, $p_x$ represents the \ac{pdf} of $x$, $p_{x|y}$ represents the conditional \ac{pdf} of $x$ given $y$, and $p_{xy}$ represents the joint \ac{pdf} of $x$ and $y$.
If $\mathbold{a} \in \mathbb{C} ^n$ and $\mathbold{b} \in \mathbb{C} ^m $ are random vectors, $\mathbold{K_{ab}} = \mathbb{E}[\mathbold{ab}^\mathrm{H}]$ denotes their $n \times m$ covariance matrix.  
Finally, $\log$ denotes the natural logarithm.
\section{System Model}
\label{par:model}

We consider a constellation of $m$ \acp{sv} (Alice, block $\rm{A}$ in Fig. \ref{fig:Model}) transmitting signals to a receiver (Bob, block $\rm{B}$ in Fig. \ref{fig:Model}). By estimating the relative delay among signals received from the $m$ \acp{sv} and by knowing the position of the \acp{sv}, Bob estimates its position through ranging techniques \cite{kaplan}. A third device (Eve, block $\rm{E}$ in Fig. \ref{fig:Model}) both receives the signals from the \acp{sv} and transmits them to Bob. The aim of Bob is to authenticate the received signal, i.e., to determine whether it comes from the constellation of \acp{sv} or from Eve. 
In turn, Eve aims at transmitting a signal that can be confused as authentic by Bob but has different delays among \ac{sv} signals, to induce a different position estimation. A reliable and asynchronous side communication channel (which cannot be used for position estimation) enables the transmission of authenticated data from Alice to Bob for the verification of the \ac{gnss} signal, while Eve cannot access it to build the attack. A possible way to implement this channel is through delayed authentication techniques \cite{delayauth}.
In this paper, we investigate how Alice can transmit its signals (on both the main and the side channel) and how Bob can perform the verification step of the authentication procedure. Also, we study possible attack strategies by Eve.

\subsection{Transmission Procedure}
The $i$-th \ac{sv}, $i \in \{1, 2, ..., m\}$, broadcasts a radio signal represented by its discrete-time complex baseband equivalent vector $\bar{\mathbold{x}}_i \in \mathbb{C} ^n$, with $\bar{\mathbold{x}}_i$ independent from $\bar{\mathbold{x}}_j$ if $i \neq j$. 
We define the \textit{transmitted word} $\mathbold{x}$ as the concatenation of the signals transmitted by all \acp{sv}: $\mathbold{x}=[\bar{\mathbold{x}}_1, \bar{\mathbold{x}}_2, ..., \bar{\mathbold{x}}_m] \, \in \mathbb{C}^{mn}$; word $\mathbold{x}$ is also delivered to Bob over the side channel.
It is important to note that the side information provided to the receiver can also be a compressed lossless version of the transmitted word $\mathbold{x}$. 
Two delays are associated to each signal $\bar{\mathbold{x}}_i$, namely $\tau_{\mathrm{B},i}$ and $\tau_{\mathrm{E},i}$, indicating the propagation time of $\bar{\mathbold{x}}_i$ from the $i$-th \ac{sv} to Bob and to Eve, respectively. 
Without loss of generality we assume $\min_i \{\tau_{\mathrm{B},i}\} =\min_i \{\tau_{\mathrm{E},i}\}= 0$, so that vectors $\mathbold{\tau}_\mathrm{B} = [\tau_{\mathrm{B},1}, \tau_{\mathrm{B},2}, ..., \tau_{\mathrm{B},m}] $ and $\mathbold{\tau}_\mathrm{E} = [\tau_{\mathrm{E},1}, \tau_{\mathrm{E},2}, ..., \tau_{\mathrm{E},m}]$ collect the relative delays between the signals coming from the $m$ satellites.
Moreover, we define $\delta_\mathrm{B} = \max_i  \{\tau_{\mathrm{B},i}\}$ and $\delta_\mathrm{E} = \max_i  \{\tau_{\mathrm{E},i}\}$.

Signal $\bar{\mathbold{x}}_i$, $i \in \{1, 2, ..., m\}$, is transmitted through two linear channels $\mathbold{y}_i=\bar{\mathbold{A}_i}\bar{\mathbold{x}_i}$ and $\mathbold{z}_i=\bar{\mathbold{F}_i}\bar{\mathbold{x}_i}$, providing the useful signals received by Bob and Eve, respectively.
Matrices $\bar{\mathbold{A}}_i$ and $\bar{\mathbold{F}}_i$ are determined by relative delays between signals, the fading environment, fluctuations in atmospheric parameters, signal distortion, and channel gains. Moreover, $\bar{\mathbold{A}}_i$ and $\bar{\mathbold{F}}_i$ include proper padding to guarantee that each channel output vector has the same size.
We denote the concatenation of matrices $\bar{\mathbold{A}}_i$ and $\bar{\mathbold{F}}_i$, for $i=1,...,m$, as $\mathbold{A} = [\bar{\mathbold{A}}_1, \bar{\mathbold{A}}_2, ..., \bar{\mathbold{A}}_m] \in \mathbb{C}^{(n+\delta_\mathrm{B}) \times (mn)}$ and $\mathbold{F}= [\bar{\mathbold{F}}_1, \bar{\mathbold{F}}_2, ..., \bar{\mathbold{F}}_m] \in \mathbb{C}^{(n+\delta_\mathrm{E}) \times (mn)}$.

In nominal conditions, Bob receives the sum of the signals coming from the $m$ satellites, 
\begin{equation}
\mathbold{y}_0 = \sum_{i=1}^{m}\mathbold{y}_i = \mathbold{Ax}\,.
\end{equation}
The same holds for Eve, who obtains 
\begin{equation}
\mathbold{z}_0 = \sum_{i=1}^{m}\mathbold{z}_i = \mathbold{Fx}\,.
\end{equation}
Furthermore, $\mathbold{y}_0$ and $\mathbold{z}_0$ are corrupted by \ac{awgn}, represented by two circularly symmetric complex Gaussian random vectors with independent entries $\mathbold{\omega}_\mathrm{B} \sim \mathcal{CN}(\mathbold{0}, \mathbold{K}_\mathrm{B})$ and $\mathbold{\omega_\mathrm{E}} \sim \mathcal{CN}(\mathbold{0}, \mathbold{K}_\mathrm{E})$, where $\mathbold{K}_\mathrm{B} =2\sigma^2_\mathrm{B} \mathbold{I}_{n+\delta_\mathrm{B}}$, $\mathbold{K}_\mathrm{E} = 2\sigma^2_\mathrm{E} \mathbold{I}_{n+\delta_\mathrm{E}}$, and
$\sigma^2_\mathrm{B}$, $\sigma^2_\mathrm{E}$ are the variances of each component (real or imaginary) of the complex noise at Bob's and Eve's receiver, respectively. 
Then, the signals received in nominal conditions by Bob and Eve are, respectively,
\begin{equation}
\mathbold{y}= \mathbold{y}_0+\mathbold{\omega_\mathrm{B}}\,,\;\;\;
\mathbold{z}= \mathbold{z}_0+\mathbold{\omega_\mathrm{E}}\,.
\end{equation}   


\subsection{Attack Strategy}

The goal of Eve is to falsify the propagation times thus introducing the forged relative delays $\mathbold{\tau}_\mathrm{f} = [\tau_{\mathrm{f},1}, \tau_{\mathrm{f},2}, ..., \tau_{\mathrm{f},m}]$ corresponding to a possibly different position than the actual receiver location. 
We assume that: (i) Eve does not know $\mathbold{x}$ but only knows $\mathbold{z}$, which is a noisy and reduced-size version of $\mathbold{x}$; (ii) Eve knows the joint distributions of $\mathbold{x}$, $\mathbold{y}$, and $\mathbold{z}$; (iii) Eve knows the channels  matrices of links $\rm{A} \rightarrow \rm{B}$, $\rm{A} \rightarrow \rm{E}$, and $\rm{E} \rightarrow \rm{B}$, and the corresponding noise statistics. We remark that the attacker cannot process each satellite signal separately: such processing would in fact require the knowledge of each word $\bar{\mathbold{x}}_i$, $i=1,\hdots,m$.

We assume that Eve can directly design the vector $\mathbold{v_0}$ received by Bob when under attack, apart from the receiver noise. Thus, when under attack the signal received by Bob is
\begin{equation}
\mathbold{v} = \mathbold{v_0}+\mathbold{\omega_\mathrm{B}}\,.
\end{equation}
where $\mathbold{v}$, $\mathbold{v}_0 \in \mathbb{C}^{n+\delta_\mathrm{f}}$, with $\delta_\mathrm{f} = \max \mathbold{\tau}_\mathbold{f}$.

Eve’s spoofing strategy can exploit the information carried by her observations $\mathbold{z}$ and, for the sake of generality, we consider that Eve adopts a probabilistic strategy, characterized by the conditional pdf $p_{\mathbold{v_0}|\mathbold{z}}$.
Moreover, we assume that Eve knows the statistics of the noise at Bob, so that the attack strategy can be described by the pdf $p_\mathbold{v|z}$. Since the observation $\mathbold{z}$ encloses all the information Eve can exploit to deceive Bob, we conclude that the forging strategy $\mathbold{v}$ is conditionally independent of the transmitted word $\mathbold{x}$, given $\mathbold{z}$.

Let the received signal by Bob be
\begin{equation}
    \mathbold{r} = \begin{cases} \mathbold{y} & \mbox{if Bob is locked on the legitimate signal } (b=0)\\
    \mathbold{v} & \mbox{if Bob is locked on the spoofing signal } (b=1) \,,\end{cases}
    \label{eqn:recsignal}
\end{equation}
where the binary variable $b$ in (\ref{eqn:recsignal}) indicates the legitimate/attack state.
We assume that Eve is able to completely cancel the signal $\mathbold{y}_0$ at Bob, thus considering the worst case scenario where Bob acquires and locks onto the spoofed signal $\mathbold{v}$ \cite{AssSpoofTh}. Therefore, Eve aims at preventing Bob from distinguishing between $\mathbold{v}$ and the legitimate $\mathbold{y}$ that would be obtained with $\mathbold{\tau}_\mathrm{B} = \mathbold{\tau}_\mathrm{f}$.

\subsection{Authentication Procedure}
The goal of the legitimate receiver Bob is to figure out if the signal $\mathbold{r}$ he receives corresponds to the authentic signal $\mathbold{y}$, or to a signal $\mathbold{v}$ that has been forged by Eve. In making this decision, Bob can make use of his knowledge of $\mathbold{x}$, which has been disclosed by Alice through the side channel.
We remark that the message transmitted through the side channel can be a lossless compressed version of $\mathbold{x}$. To detect the spoofing attack, Bob performs an authentication test, wherein, given $\mathbold{x}$ and the observation $\mathbold{r}$, Bob chooses between the two hypotheses:
\begin{align}
\mathcal{H}_0 &: \mathbold{r} = \mathbold{y}\,, \mbox{ the message is from Alice} \label{eqn:h0},\\
\mathcal{H}_1 &: \mathbold{r} = \mathbold{v}\,,  \mbox{ the message was forged} \label{eqn:h1}.
\end{align}
In Fig. \ref{fig:Model}, the correct verification is achieved when $\hat{b} = b$. The authentication procedure is summarized in block $D$, which has the received signal $\mathbold{r}$ as input and outputs the Boolean value $\hat{b}$.

It is worth noting that the model in Fig. \ref{fig:Model} includes previous models from the literature, as particular cases.
In the scheme proposed in \cite{Scott2013,chimera,chimeraION}, only one satellite has been considered, which can be cast into our model by taking $m=1$.  In these solutions, a small part of the spreading code is superimposed with a
secret, cryptographically generated sequence, which can be subsequently reproduced by the receivers when they become aware of the key, while navigation message data are protected by digitally signing most or all the data.
So, word $\mathbold{x}$ is the signal obtained by the superposition of the signed navigation data with the signed version of the spreading code, followed by a \ac{bpsk} modulation. Then, with a delay with respect to $\mathbold{x}$, the key is broadcast as side information, so that the receiver is able to reconstruct $\mathbold{x}$.
A similar approach can be adopted to describe Galileo \ac{osnma} \cite{OSNMAart,OSNMAart2}, possibly combined with \ac{cas} \cite{cas22} or \ac{acas} features \cite{acasICL}.
The model in \cite{poltronieriICL} can be cast into this frame by restricting $\mathbold{x}$ to be a binary watermarking sequence, and the attacker to have complete ignorance on it, thus, removing his observation $\mathbold{z}$.
In the authentication scheme of \cite{ANTomasin}, $\mathbold{x}$ is the superposition of the transmitted authentication message and the \ac{an} component. Then, the authentication message and the \ac{an} are transmitted to the legitimate receiver through an authenticated channel.

All the parameters presented in this Section are summarized in Table \ref{tab:param}.
\begin{table} 
    \centering
    \caption{System parameters, $i \in {1, 2, ..., m}$}\label{tab:param}
    \begin{tabularx}{8.5cm}{r|l}
\toprule
 \textbf{Symbol} & \textbf{Definition}     \\   
\midrule 
$m$ & Number of satellites  \\
$n$ & Length of the discrete time signal $\mathbold{x}_i$  \\
$\mathbold{x}_i$ & Discrete time signal transmitted by the $i$-th SV \\
$\mathbold{x}$ & Transmitted word   \\
$\tau_{\mathrm{B},i}$ & Propagation time of $\mathbold{x}_i$ from the $i$-th SV to B   \\
$\tau_{\mathrm{E},i}$  & Propagation time of $\mathbold{x}_i$ from the $i$-th SV to E   \\
$\tau_{\mathrm{f},i}$  & False propagation time related to $\mathbold{x}_i$           \\
$\mathbold{\tau_\mathrm{B}}$  & Vector of propagation times from A to B           \\
$\mathbold{\tau_\mathrm{E}}$  & Vector of propagation times from A to E           \\
$\mathbold{\tau_\mathrm{f}}$  & Vector of false propagation times introduced by E   \\
$\delta_\mathrm{B}$   & Largest propagation time of signals traveling from A to B   \\
$\delta_\mathrm{E}$   & Largest propagation time of signals traveling from A to E   \\
$\delta_\mathrm{f}$   & Largest propagation time of signals traveling from E to B     \\
$\mathbold{A}$  & A$\rightarrow$ B channel matrix \\
$\mathbold{F}$  & A$\rightarrow$ E channel matrix \\
$\mathbold{\omega_\mathrm{B}}$  & AWGN in the channel A$\rightarrow$ B and E$\rightarrow$ B    \\
$\mathbold{\omega_\mathrm{E}}$  & AWGN in the channel A$\rightarrow$ E     \\
$\sigma^2_\mathrm{B}$  & Noise variance per component channel A$\rightarrow$ B         \\
$\sigma^2_\mathrm{E}$  & Noise variance per component channel A$\rightarrow$ E                \\
$\mathbold{K}_\mathrm{B}$  & Covariance matrix of $\mathbold{\omega_\mathrm{B}}$             \\
$\mathbold{K}_\mathrm{E}$  & Covariance matrix of $\mathbold{\omega_\mathrm{E}}$             \\
$\mathbold{y}$  & Received signal through the AWGN channel A$\rightarrow$ B      \\
$\mathbold{z}$  & Received signal through the AWGN channel A$\rightarrow$ E     \\
$\mathbold{v_0}$  & Signal generated by E                        \\
$\mathbold{v}$   &  Received signal through the AWGN channel E$\rightarrow$ B \\
$\mathbold{r}$   & Signal received by B                                      \\
$\hat{b}$  & Decision made by the receiver B                                 \\
$b$        & Correct value for $\hat{b}$    \\                                                            
\bottomrule
\end{tabularx}
\end{table}

\subsection{Performance Metric}
\label{par:problem}
The performance of an authentication system is assessed by: \textit{a)} the type-\RN{1} (\textit{false alarm}) error probability $p_\mathrm{FA}$, i.e., the probability that Bob discards a message as forged by Eve while it is coming from Alice; \textit{b)} the type-\RN{2} (\textit{missed detection}) error probability $p_\mathrm{MD}$, i.e., the probability that Bob accepts a message coming from Eve as legitimate.
The \ac{lrt} is the optimal detection method that minimizes the false alarm probability for a fixed missed detection given $p_\mathbold{x}$, and $p_\mathbold{v|z}$ \cite{signalProcKay}. However, in general, the analytical derivation of these \ac{pdf} is hardly feasible.
Therefore, theoretical bounds on the achievable error probability region are useful to establish the effectiveness of practical schemes.

A first bound on the achievable error region, which is the set of achievable points in the $(p_\mathrm{FA}, p_\mathrm{MD})$ plane,  for a given attack strategy is given by  the \ac{kl} divergence. 
In fact, from \cite{MaurerAuthTh} and \cite{CachinSteg} we have
\begin{equation}
\mathbb{D}(p_{\hat{b}|\mathcal{H}_1} \| p_{\hat{b}|\mathcal{H}_0}) \leq
\mathbb{D}(p_{\mathbold{rx}|\mathcal{H}_1} \|  p_{\mathbold{rx}|\mathcal{H}_0}) = \mathbb{D}(p_{\mathbold{xv}}\| p_{\mathbold{xy}}).
\label{eqn:d1}
\end{equation}
In (\ref{eqn:d1}) we have considered the joint \ac{pdf} $p_\mathbold{rx}$ since we suppose that, at the time of verification, the legitimate receiver knows $\mathbold{x}$, and the decision $\hat{b}$ is taken based on both inputs. Therefore, defining the function
\begin{equation}
h(p_\mathrm{MD},p_\mathrm{FA}) \triangleq p_\mathrm{MD} \log \frac{p_\mathrm{MD}}{1-p_\mathrm{FA}} + (1-p_\mathrm{MD})\log \frac{1-p_\mathrm{MD}}{p_\mathrm{FA}}\,,
\label{eqn:h}
\end{equation}
with $p_\mathrm{FA}, p_\mathrm{MD} \in [0,1]$, and observing that $p_{\hat{b}|\mathcal{H}_0}(1) = p_\mathrm{FA}$, $p_{\hat{b}|\mathcal{H}_0}(0) = 1-p_\mathrm{FA}$, 
$p_{\hat{b}|\mathcal{H}_1}(1) = 1-p_\mathrm{MD}$ and $p_{\hat{b}|\mathcal{H}_1}(0) = p_\mathrm{MD}$, (\ref{eqn:d1}) can be rewritten as
\begin{equation}
h(p_\mathrm{MD},p_\mathrm{FA}) \leq \mathbb{D}(p_{\mathbold{xv}}\| p_{\mathbold{xy}}).
\label{eqn:d2}
\end{equation}
This limits the region of achievable $(p_\mathrm{FA}, p_\mathrm{MD})$ values, depending on $\mathbb{D}(p_{\mathbold{xv}}\| p_{\mathbold{xy}})$, for any decision mechanism choice.~\footnote{The bound in \eqref{eqn:d2} is tight when Bob knows Eve's attack strategy and Eve does not know Bob's defense strategy, however, the bound also holds in the case of interest where Bob does not know what Eve does and Eve knows what Bob does.}

On one hand, the aim of the attacker Eve is to narrow the achievable region, by making the value of $\mathbb{D}(p_{\mathbold{xv}}\|p_{\mathbold{xy}})$ as small as possible, operating on the attack strategy $p_\mathbold{v|z}$. On the other hand, Alice aims at enlarging the achievable region, by properly choosing the distribution of the transmitted word $\mathbold{x}$ in order to increase $\mathbb{D}(p_{\mathbold{xv}}\|p_{\mathbold{xy}})$. Therefore, the defense strategy is defined by the \ac{pdf} $p_\mathbold{x}$.

The metric $\mathbb{D}(p_{\mathbold{xv}}\| p_{\mathbold{xy}})$ can be expressed in terms of attack and defense strategies as
\begin{multline}
    \mathbb{D}(p_{\mathbold{xv}}\| p_{\mathbold{xy}}) = \iint p_\mathbold{x}(a)p_\mathbold{v|x}(b|a)\log\frac{p_\mathbold{v|x}(b|a)}{p_\mathbold{y|x}(b|a)}\mbox{d}a\,\mbox{d}b\\ = \iint p_\mathbold{x}(a) \int p_\mathbold{v|z}(b|c)p_\mathbold{z|x}(c|a)\,\mbox{d}c \\
    \;\;\;\; \cdot\log\frac{\int p_\mathbold{v|z}(b|c)p_\mathbold{z|x}(c|a)\mbox{d}c}{p_\mathbold{y|x}(b|a)}\,\mbox{d}a\,\mbox{d}b   \,.
\end{multline}
Highlighting the contribution of attack and defense strategies, let us define
\begin{equation}
    f(p_\mathbold{x}, p_\mathbold{v|z}) \triangleq \mathbb{D}(p_{\mathbold{xv}}\|p_{\mathbold{xy}})\,,
    \label{eqn:f}
\end{equation}
with fixed channels $p_\mathbold{y|x}$ and $p_\mathbold{z|x}$.
So, the task that we will address in this paper from the point of view of the defense can be defined as the following maximin problem:
\begin{equation}
\max_{p_\mathbold{x}}\,  \min_{p_{\mathbold{v}|\mathbold{z}}} f(p_\mathbold{x}, p_\mathbold{v|z}).
\label{eqn:maxmin}
\end{equation}

An important difference that distinguishes attack and defense strategies is that Eve knows the value of $\mathbold{\tau_\mathrm{E}}$ and the victim position exactly, whereas Alice's defense strategy must be robust and symmetrical with respect to all potential receiver and channel realizations, described by matrices $\mathbold{A}$ and $\mathbold{F}$.

In Section \ref{par:min} we will first address the inner minimization in (\ref{eqn:maxmin}), considering the model presented in Section \ref{par:model} and assuming that $\mathbold{x}$ is Gaussian distributed. Then, after discussing the achievable \ac{kl} divergence values, the maximization task in (\ref{eqn:maxmin}) will be finally investigated in Section \ref{par:defense}.

\section{Attack Strategy}
\label{par:min}
We now focus on the attack strategy optimization, i.e., from (\ref{eqn:maxmin}), we aim at deriving the conditional \ac{pdf} $p_\mathbold{v|z}$ such that
\begin{equation}
p_{\mathbold{v}|\mathbold{z}}^\star  = \argmin_{p_{\mathbold{v}|\mathbold{z}}} f(p_\mathbold{x}, p_\mathbold{v|z})\,.
\label{eqn:argmin}
\end{equation}

We assume that $\mathbold{x}$ is a zero mean Gaussian random vector, so $p_\mathbold{x} \sim \mathcal{N}(0, \mathbold{K_x})$ and the optimality of this choice will be proven in Section \ref{par:defensetx}.
Under this assumption, as proved in \cite{Ferrante}, there exists an optimal attack $p_\mathbold{v|z}^\star$ minimizing the divergence in (\ref{eqn:f}).
\begin{theorem}\label{th:1}
Given the zero mean, jointly Gaussian random vectors $\mathbold{x}$,$\mathbold{y}$, and $\mathbold{z}$, the optimal attack $p_\mathbold{v|z}^\star$ minimizing the divergence in (\ref{eqn:f}), with fixed channels $p_\mathbold{y|x}$ and $p_\mathbold{z|x}$, under the constraint that the random vectors $\mathbold{v}$ and $\mathbold{x}$ are conditionally independent given $\mathbold{z}$, belongs to the class 
\begin{equation}
\mathcal{C} = \left\{ p_\mathbold{v|z} \sim \mathcal{N}( \mathbold{Gz},\,\mathbold{CC^\mathrm{H}} + {\mathbold{K_\mathrm{B}}})\right\}\,,
\label{eqn:classCv}
\end{equation}
and can therefore be written as
a linear transformation of $\mathbold{z}$, plus independent additive white complex Gaussian noise, i.e.,
\begin{equation}
\mathbold{v} = \mathbold{Gz}+\mathbold{C\omega_\mathrm{c}}+\mathbold{\omega_\mathrm{B}},
\label{eqn:classv}
\end{equation}
where $\mathbold{G} \in \mathbb{C}^{(n+\delta_\mathrm{f}) \times (n+\delta_\mathrm{E})}$, $\mathbold{C} \in \mathbb{C}^{(n+\delta_\mathrm{f}) \times (n+\delta_\mathrm{f})}$. 
\end{theorem}

\begin{proof}
See \cite[Theorem 2]{Ferrante}.
\end{proof}

In particular, the \ac{pdf} $p_{\mathbold{v}|\mathbold{z}}^\star $ that solves (\ref{eqn:argmin}) is computed by optimizing over $\mathbold{G}$ and $\mathbold{C}$, so (\ref{eqn:argmin}) becomes
\begin{equation}
(\mathbold{G^\star }, \mathbold{C^\star }) = \argmin_\mathbold{G, C} f(p_\mathbold{x}, p_\mathbold{v|z})\,.
\label{eqn:argminGC}
\end{equation}
The variance and covariance matrices of the signals defined so far are given by
\begin{subequations}
\begin{align}
    \mathbold{K_{xy}} &= \mathbold{K_xA}^\mathrm{H}\,, \label{eqn:Kxy}\\
    \mathbold{K_y} &= \mathbold{AK_xA}^\mathrm{H} +  \mathbold{K_\mathrm{B}} \label{eqn:Ky}\,,\\
    \mathbold{K_{xz}} &= \mathbold{K_xF}^\mathrm{H}  \label{eqn:Kxz}\,,\\
    \mathbold{K_z} &=\mathbold{FK_xF}^\mathrm{H} +  \mathbold{K_\mathrm{E}} \,, \label{eqn:Kz}\\
    \mathbold{K_{xv}} &= \mathbold{K_xF}^\mathrm{H}\mathbold{G}^\mathrm{H} \,, \label{eqn:Kxv}\\
    \mathbold{K_v} &= \mathbold{GFK_xF}^\mathrm{H}\mathbold{G}^\mathrm{H} +  \mathbold{GK_\mathrm{E}G}^\mathrm{H} + \mathbold{CC}^\mathrm{H} +{\mathbold{K_\mathrm{B}}}\,,\label{eqn:Kv}
\end{align}
\label{eqn:covmat}
\end{subequations}
Considering \eqref{eqn:covmat} and following the derivations of \cite{Ferrante}, 
the optimal matrices are 
\begin{equation}
      \mathbold{G^\star } =  \mathbold{AK_xF}^\mathrm{H}(\mathbold{FK_xF}^\mathrm{H})^\dagger\,, 
      \label{eqn:Gstar}\\
\end{equation}
and $\mathbold{C}^\star \in \mathbb{C}^{(n+\delta_\mathrm{f}) \times (n+\delta_\mathrm{f})}$ is the square root of the covariance matrix $\mathbold{K_{v_0|z}}$ of the signal $\mathbold{v}_0$ given $\mathbold{z}$. As outlined in \cite{Ferrante}, $\mathbold{C}^\star$ is obtained is obtained either with a closed form expression
or through an iterative process.


\subsection{K-L Divergence for Attacks in $\mathcal{C}$}
\label{par:KLdivwithv}
In this Section, we will derive an analytical expression for $f(p_\mathbold{x}, p_\mathbold{v|z})$ when $p_\mathbold{x}$  is a generic \ac{pdf} with covariance matrix $\mathbold{K_x}$.
Under the assumption that $p_\mathbold{v|z} \in \mathcal{C}$, defining $\mathbold{B} \triangleq \mathbold{GF}$ and $\mathbold{\eta} \triangleq \mathbold{G\omega_\mathrm{E}}+\mathbold{C\omega_\mathrm{c}}+{\mathbold{\omega_\mathrm{B}}}$, $\mathbold{v}$ can be written as (see (\ref{eqn:classv}))
\begin{equation}
\mathbold{v} = \mathbold{Bx}+\mathbold{\eta}\,,
\label{eqn:v2}
\end{equation}
with 
\begin{equation}
    \mathbold{\eta} \sim  \mathcal{N}(\mathbold{0}, \mathbold{K_{\eta}})\;,\;\;\mathbold{K_{\eta}} = \mathbold{GK_\mathrm{E}G}^\mathrm{H}+\mathbold{CC}^\mathrm{H}+{\mathbold{K_\mathrm{B}}}\,. \label{eqn:Keta}
\end{equation}


Considering the definition of $\mathbold{v}$ given in (\ref{eqn:v2}), the metric $f(p_\mathbold{x},p_\mathbold{v|z})$ can be computed for a generic distribution $p_\mathbold{x}$ and $p_\mathbold{v|z}\in\mathcal{C}$ as
\begin{align}
\begin{split}
f(p_\mathbold{x}&, p_\mathbold{v|z}) =\,\mathbb{E} \left[ \log \frac{p_{\mathbold{xv}}(\mathbold{x,v})}{p_{\mathbold{xy}}(\mathbold{x,v})} \right] = \mathbb{E} \left[ \log \frac{p_{\mathbold{v|x}}(\mathbold{v|x})}{p_{\mathbold{y|x}}(\mathbold{v|x})} \right] \\
=&\, \mathbb{E} \left[ \log \frac{p_{\mathbold{\eta}}(\mathbold{v-Bx})}{p_{\mathbold{\omega_\mathrm{B}}}(\mathbold{v-Ax})} \right]   
=\, \mathbb{E} \left[ \log\frac{p_{\mathbold{\eta}}(\mathbold{\eta})}{p_{\mathbold{\omega_\mathrm{B}}}(\mathbold{\eta}-\mathbold{(A-B)x})}\right]\\
=&\, \mathbb{E} \left[ \log\frac{p_{\mathbold{\eta}}(\mathbold{\eta})}{p_{\mathbold{\omega_\mathrm{B}}}(\mathbold{\eta})}\right] + \,\mathbb{E} \left[\log\frac{p_{\mathbold{\omega_\mathrm{B}}}(\mathbold{\eta})}{p_{\mathbold{\omega_\mathrm{B}}}(\mathbold{\eta-(A-B)x})} \right]  \\
=&\, \mathbb{D}(p_\mathbold{\eta}\| p_\mathbold{{\omega_\mathrm{B}}}) + \mathbb{E}\left[ ((\mathbold{A-B})x)^\mathrm{H} \mathbold{K_\mathrm{B}}^{-1}(\mathbold{A-B})x\right]\\
=&\, \frac{1}{2}\log\frac{|\mathbold{K_\mathrm{B}}|}{|\mathbold{K_{\eta}}|} +  \frac{1}{2}\tr(\mathbold{K_{\eta}K_\mathrm{B}}^{-1}) - \frac{(n+\delta_\mathrm{f})}{2} \\ &+ \frac{1}{2}\left[\tr\left((\mathbold{B-A})\mathbold{K_x}(\mathbold{B-A})^\mathrm{H}\mathbold{K_\mathrm{B}}^{-1} \right)  \right]\\
=&\, \frac{1}{2} \left[ \sum_{i=1}^{n+\delta_\mathrm{f}}\left(\frac{\lambda_i}{\sigma^2_\mathrm{B}}-\log\left(\frac{\lambda_i}{\sigma^2_\mathrm{B}} \right)\right) -(n+\delta_\mathrm{f}) \right] \\ &+ \frac{1}{2\sigma^2_\mathrm{B}}\tr\left((\mathbold{B-A})\mathbold{K_x}(\mathbold{B-A})^\mathrm{H}\right)\,,
\end{split} \label{eqn:divkx} 
\end{align}
where $\lambda_i$, $i \in \{1, 2, ..., n+\delta_\mathrm{f}\}$, are the eigenvalues of $\mathbold{K_\eta}$. We remark that (\ref{eqn:divkx}) holds for any distribution $p_\mathbold{x}$, as long as the attack \ac{pdf} $p_\mathbold{v|z}$ is taken from $\mathcal{C}$.

For convenience, we analyze the two terms of \eqref{eqn:divkx} separately
\begin{align}
   t_1 &= \frac{1}{2} \left[ \sum_{i=1}^{n+\delta_\mathrm{f}}\left(\frac{\lambda_i}{\sigma^2_\mathrm{B}}-\log\left(\frac{\lambda_i}{\sigma^2_\mathrm{B}} \right)\right) -(n+\delta_\mathrm{f}) \right], \label{eqn:t1}\\
    t_2 &= \frac{1}{2\sigma^2_\mathrm{B}}\tr\left((\mathbold{B-A})\mathbold{K_x}(\mathbold{B-A})^\mathrm{H}\right)\,. \label{eqn:t2}
\end{align}
Since $c - 1 \geq \log c, \; \forall c \in \mathbb{R}^+$, we can state that each term of the sum in $t_1$ is non-negative. Moreover, $t_1=0$ if and only if  $\lambda_i / \sigma^2_\mathrm{B} = 1, \; \forall i \in \{ 1, 2, ..., n+\delta_\mathrm{f}\}$, i.e., if and only if the attacker manages to construct $\mathbold{K_\eta} = \mathbold{K_\mathrm{B}}$. 
On the other hand, the term $t_2$ in \eqref{eqn:divkx} is independent of the attacker noise $\mathbold{\omega_\mathrm{E}}$, because it does not depend on $\mathbold{K_\eta}$, while it depends on the covariance matrix $\mathbold{K_x}$, the legitimate receiver noise power $\sigma^2_\mathrm{B}$ in the nominal case, and the difference $\mathbold{A}-\mathbold{B}$, where $\mathbold{A}$ and $\mathbold{B}$ are the authentic and the forged channel matrix, respectively. Therefore, the following inequality holds
\begin{equation}
   f(p_\mathbold{x}, p_\mathbold{v|z}) \geq t_2\,,\,\; \forall p_\mathbold{v|z}\in\mathcal{C}\,.
   \label{eqn:divbound}
\end{equation}

\subsection{K-L Divergence Under Optimal Attack}
From (\ref{eqn:divkx}) and (\ref{eqn:divbound}), we observe that, for a generic $p_\mathbold{x}$ with covariance matrix $\mathbold{K_x}$, once the values of $\mathbold{A}$, $\mathbold{K_x}$, and $\sigma^2_\mathrm{B}$ are fixed, the lower bound for the \ac{kl} divergence can be expressed as a function of $\mathbold{B}$ as 
\begin{equation}
\mathbb{D}_{min}(\mathbold{B})  \triangleq  \frac{1}{2\sigma^2_\mathrm{B}}\tr\left((\mathbold{B-A})\mathbold{K_x}(\mathbold{B-A})^\mathrm{H}\right)\,.
\label{eqn:dmin}
\end{equation}
In particular. when the attacker succeeds in constructing $\mathbold{K_{\eta}}=\mathbold{K_\mathrm{B}}$, then $t_1=0$, and
\begin{equation}
f(p_\mathbold{x}, p_\mathbold{v|z}^\star) = \mathbb{D}_{min}(\mathbold{B}^\star) \label{eqn:dmin*}\,.
\end{equation}

A worth noting consideration is that, for a fixed $\mathbold{K_x}$, $p_{\mathbold{v}|\mathbold{z}}^\star$ achieves the minimum \ac{kl} divergence among all the possible attack strategies $p_\mathbold{v|z}\in\mathcal{C}$, regardless of the shape of $p_\mathbold{x}$, thus
\begin{equation}
f(p_\mathbold{x}, p_\mathbold{v|z}) \geq f(p_\mathbold{x}, p_\mathbold{v|z}^\star)\,,\;\; \forall \; p_{\mathbold{v}|\mathbold{z}} \in \mathcal{C}. 
\end{equation}


In particular, when $\mathbold{K_x} = M_x\mathbold{I}_{mn}$ (as further discussed in Section \ref{par:defensetx}), $\mathbb{D}_{min}(\mathbold{B})$ can be written as
\begin{equation}
    \mathbb{D}_{min}(\mathbold{B}) = \frac{M_x}{2 \sigma^2_\mathrm{B}} \; \|\mathbold{A}-\mathbold{B}\|_F^2 \,  = \,\frac{mn}{2} \, k \,\Lambda_\mathrm{AB}\,,
    \label{eqn:div3terms}
\end{equation}
where 
\begin{equation}
    k = \frac{\|\mathbold{A}-\mathbold{B}\|_F^2}{\|\mathbold{A}\|^2_F}
\end{equation}
represents a diversity index between $\mathbold{A}$ and $\mathbold{B}$, and
\begin{equation}
    \Lambda_\mathrm{AB} = \frac{M_x}{\sigma^2_\mathrm{B}} \, \frac{\|\mathbold{A}\|^2_F}{mn}\,,  \label{eqn:lambda0} 
\end{equation}
represents the average received \ac{snr} at Bob.
It is worth noting that the attacker can always choose $\mathbold{G}$ in such a way that $k\leq1$ (trivially setting $\mathbold{G} = 0$ we have $k = 1$). 
Moreover, in (\ref{eqn:lambda0}), the term 
\begin{equation}
    \frac{\|\mathbold{A}\|_F^2}{mn} = \frac{1}{m} \sum^{m}_{i=1} \frac{1}{n} \|\mathbold{A}\|_F^2
\end{equation}
 represents the average energy of the $m$ impulsive responses of the legitimate channels.
 Therefore, (\ref{eqn:div3terms}) describes $\mathbb{D}_{min}(\mathbold{B})$ in terms of the total length of the $m$ transmitted signals ($mn$), a measure of the difference between the channel matrices ($k$), where $\mathbold{A}$ is the matrix of the legitimate channel and $\mathbold{B}$ is the matrix of the forged channel, the \ac{snr} of the legitimate channel $A\rightarrow B$ ($\Lambda_\mathrm{AB}$).

\color{black}
\subsection{Limiting Scenarios}
We now discuss some limiting scenarios, in which the considered spoofing attack  achieves complete indistinguishability from the legitimate signal and hence cannot be detected:
\begin{description}
    \item[\textbf{S1:}] $\mathbold{A} = \mathbold{F}$, that is the case wherein Eve performs a meaconing attack, inducing her own position onto Bob;
    \item[\textbf{S2:}] $m = 1$, so both Eve and Bob receive the signal from only one satellite, and there is not sufficient diversity;
    \item[\textbf{S3:}] the channel $\rm{A}\rightarrow\rm{B}$ is stochastically degraded with respect to the channel $\rm{A}\rightarrow\rm{E}$.
\end{description}
In the following analysis, we assume that the attacker Eve makes use of the optimal attacking strategy ${p_{\mathbold{v|z}}^\star} $.

In scenario \textbf{S1}, where $\mathbold{A}=\mathbold{F}$, we have that $\delta_\mathrm{E} = \delta_\mathrm{B} = \delta_\mathrm{f} = \delta$. From (\ref{eqn:Gstar}), Eve gets $\mathbold{G}^\star  =  \mathbold{I_{n+\delta}}$ so that $\mathbold{B}^\star  = \mathbold{G^\star F}= \mathbold{F}$, and  $\mathbold{B}^\star =\mathbold{F} = \mathbold{A}$, which implies $\mathbb{D}_{min}(\mathbold{B}^\star ) = 0$. 
Thus, the meaconing attack cannot be detected in this model.

In scenario \textbf{S2} we are supposing $m=1$. This implies that $\delta_\mathrm{E} = \delta_\mathrm{B} = \delta_\mathrm{f} =0$, so $\mathbold{A}$ and $\mathbold{F}$ are left invertible. Therefore, when $\mathbold{G^\star}$ is computed as in (\ref{eqn:Gstar}), Eve will get $\mathbold{B^\star} = \mathbold{G^\star F} = \mathbold{A}$, which implies $\mathbb{D}_{min}(\mathbold{B}^\star ) = 0$. 
Therefore, to have $\mathbb{D}_{min}(\mathbold{B^\star}) > 0$, Bob has to combine signals from $m>1$ satellites.
However, this is also a necessary condition for the \ac{gnss} receiver to calculate the \ac{pvt} solution.

In scenario \textbf{S3} we have that the channel $\rm{A}\rightarrow \rm{B}$, represented by the conditional \ac{pdf} $p_\mathbold{y|x}$, can be decomposed as the cascade of $p_\mathbold{z|x}$ and some properly chosen $p_\mathbold{y|z'}$. Therefore, in this case Eve can choose $p_\mathbold{v|z} = p_\mathbold{y|z'}$ to obtain $p_\mathbold{y|x}=p_\mathbold{v|x}$.
Moreover, we have $\mathbold{y}=\mathbold{G'z'}+\mathbold{C'\omega_\mathrm{c}}$, and Eve chooses $\mathbold{G^\star}=\mathbold{G'}$ and $\mathbold{C^\star} = \mathbold{C'}$, so that $\mathbold{B^\star} = \mathbold{A}$ and $\mathbb{D}_{min}(\mathbold{B}^\star) = 0$. Therefore, in this case, the attack goes undetected.

The more general, and more realistic, spoofing scenario occurs when $m>1$, $\mathbold{A}\neq\mathbold{F}$ and $\sigma_\mathrm{E}^2 > 0$. Moreover, the hypothesis in scenario \textbf{S3} is very pessimistic. Therefore, in a realistic scenario, with the additional assumption that $\ker(A) \nsubseteq \ker(F)$, it is always assured that $\mathbb{D}_{min}(\mathbold{B})>0$.

\section{Defense Strategy Design}
\label{par:defense}
The transmission and the attack detection strategies together determine the defense strategy.
Therefore, in this Section, we investigate how Alice designs the transmitted signal and how Bob performs the verification step of the authentication procedure. 
\subsection{Gaussian Transmission}
\label{par:defensetx}
The optimal transmission strategy $p_\mathbold{x}^\star$, is given by the maximin solution of (\ref{eqn:maxmin}). 
 We start by identifying the optimal distribution for a constrained covariance matrix, introducing the following theorem.
\begin{theorem}\label{th:NE}
Given a covariance matrix $\mathbold{K_x}$, let us define $p_\mathbold{x}^\star \sim \mathcal{N}(0, \mathbold{K_x})$. If the transmission distribution $p_\mathbold{x}$ is to be chosen among all those with zero mean and covariance $\mathbold{K_x}$, the pair of strategies $(p_\mathbold{x}^\star, p_\mathbold{v|z}^\star)$ constitutes a saddle point of the function $f(p_\mathbold{x}, p_\mathbold{v|z})$.
\end{theorem}
\begin{proof}
For any attack strategy $p_\mathbold{v|z} \in \mathcal{C}$, we can compute $f(p_\mathbold{x},p_\mathbold{v|z})$ for a generic distribution $p_\mathbold{x}$ from (\ref{eqn:divkx}). We note that, when $p_\mathbold{v|z} \in \mathcal{C}$, the \ac{kl} divergence depends on $p_{\mathbold{x}}$ only through the covariance matrix $\mathbold{K_x}$. Consequently, if $p_\mathbold{v|z} \in \mathcal{C}$, once matrices $\mathbold{A}$, $\mathbold{B}$, and $\mathbold{K_x}$ are set, then $f(p_\mathbold{x}, p_\mathbold{v|z})$ is constant for each probability distribution $p_\mathbold{x}$ chosen by Alice. Therefore, we conclude that the set of strategies ($p_\mathbold{x}^\star, p_\mathbold{v|z}^\star$) constitutes a saddle point for $f(p_\mathbold{x}, p_\mathbold{v|z})$, since neither the attacker nor the defender can gain by an unilateral change of strategy if the strategy of the other remains unchanged. In particular
\begin{align}
    f(p_\mathbold{x}, p_\mathbold{v|z}^\star) &= f(p_\mathbold{x}^\star, p_\mathbold{v|z}^\star)\,,\,\;\forall\, p_\mathbold{x}\,, \label{eqn:proof1}   \\
    f(p_\mathbold{x}^\star, p_\mathbold{v|z}) &> f(p_\mathbold{x}^\star, p_\mathbold{v|z}^\star)\,,\,\;\forall\, p_\mathbold{v|z}\,,\label{eqn:proof2}  
\end{align}
where \eqref{eqn:proof1} follows from \eqref{eqn:divkx}, when the covariance matrix $\mathbold{K_x}$ is fixed, while \eqref{eqn:proof2} holds for the optimality of $p_\mathbold{v|z}^\star$ when the transmission distribution is $p_\mathbold{x}^\star$, as stated in Theorem~\ref{th:1}.
\end{proof}

 The maximin problem in \eqref{eqn:maxmin} can be seen as a zero-sum game with two players, where $p_\mathbold{x}$ and $p_\mathbold{v|z}$ are mixed strategies, while $f(p_\mathbold{x},p_\mathbold{v|z})$ and $- f(p_\mathbold{x},p_\mathbold{v|z})$ are the average payoffs\footnote{Player's payoffs are averaged over the mixed players' strategies and the noise distribution, $f(p_\mathbold{x}, p_\mathbold{v|z}) = \mathbb{E}\left[\log \frac{p_\mathbold{xv}(\mathbold{x,v})}{p_\mathbold{xy}(\mathbold{x,v})}\right]$.} for the defender and the attacker, respectively.
 In this case, the set of strategies ($p_\mathbold{x}^\star, p_\mathbold{v|z}^\star$), constitutes one Nash equilibrium of the game. We remark that there may be many Nash equilibria, however, for the properties of zero-sum games, they all have the same average payoff \cite{libroGT}. 



From Theorem \ref{th:NE}, we conclude that the optimal defense strategy $p_\mathbold{x}^\star$ solving (\ref{eqn:maxmin}) must be a zero mean Gaussian distribution. Furthermore, Alice can choose the covariance matrix $\mathbold{K_x}$ of $p_\mathbold{x}^\star$ so that $f(p_\mathbold{x}^\star, p_\mathbold{v|z}^\star)$ is maximized while ensuring the constraint on the transmitted power $M_x$: $\tr(\mathbold{K_x}) \leq mn\,M_x$. 
Given the symmetry of the problem and the transmitter's lack of knowledge of the channel and, consequently, of the matrices $\mathbold{A}$ and $\mathbold{F}$, a reasonable choice for $\mathbold{K_x}$ is $\mathbold{K_x} = M_x \, \mathbold{I}_{mn}$.

From the theory of binary hypothesis testing, we know that if both the statistics of the legitimate and spoofed signal are known (i.e., if the victim is aware of the particular attack strategy adopted by the spoofer) the test yielding the minimum $p_\mathrm{MD}$ for any given constraint on $p_\mathrm{FA}$, is the \ac{lrt}, also known as Neyman-Pearson criterion \cite{signalProcKay,MaurerAuthTh}. Under the assumption that the attack strategy belongs to class $\mathcal{C}$, the detection problem reduces to the test
\begin{multline}
L' = (\mathbold{r}-\mathbold{Ax})^\mathrm{H}\mathbold{K_\mathrm{B}}^{-1}(\mathbold{r}-\mathbold{Ax})\\ -(\mathbold{r}-\mathbold{Bx})^\mathrm{H}\mathbold{K_{\eta}}^{-1}(\mathbold{r}-\mathbold{Bx}) \underset{H_0}{\overset{H_1}{\gtrless}} \theta\,.
\end{multline}
When both Eve and Alice play the Nash equilibrium strategies derived in Sections \ref{par:min} and \ref{par:defensetx}, Bob will use the \ac{lrt} since it is aware of the attack strategy distribution.

\subsection{Transmission Strategies With Practical Modulation Schemes}
\label{par:GaussFin}
In Section \ref{par:defensetx} we derived that the optimal defense strategy solving (\ref{eqn:maxmin}) must be a zero mean Gaussian distribution. However, in practice this is not feasible and hence we analyze the case wherein $\mathbold{x}$ has symbols from a finite-set, e.g., a QAM constellation.

In (\ref{eqn:divkx}) we showed that, when $p_\mathbold{v|z} \in \mathcal{C}$, the value of $f(p_\mathbold{x}, p_\mathbold{v|z})$ depends on $p_{\mathbold{x}}$ only through the covariance matrix $\mathbold{K_x}$. This implies that
\begin{equation}
   f(p_\mathbold{x}, p_\mathbold{v|z}) = f(p_\mathbold{x}^\star, p_\mathbold{v|z}) \,, \;\; \forall \; p_\mathbold{v|z}\in \mathcal{C} \,,
    \label{eqn:chain1}
\end{equation}
where $p_\mathbold{x}$ is a generic distribution of the signal $\mathbold{x}$, with the same covariance matrix $\mathbold{K_x}$ of $p_\mathbold{x}^\star \sim \mathcal{N}(\mathbold{0,K_x})$.
Therefore, when the attack belongs to the class $\mathcal{C}$, we may conclude that the performance in terms of divergence is the same with either Gaussian or finite-cardinality modulation.
Moreover, from (\ref{eqn:proof1}) and (\ref{eqn:proof2}) we have
\begin{equation}
    f(p_\mathbold{x}, p_\mathbold{v|z}^\star) = f(p_\mathbold{x}^\star, p_\mathbold{v|z}^\star) \leq f(p_\mathbold{x}^\star, p_\mathbold{v|z})\,, \;\; \forall \; p_\mathbold{v|z}\,.
    \label{eqn:chain3}
\end{equation}

On the other hand, $p_\mathbold{v|z}^\star$ is the optimal attack strategy only when the transmitted signal $\mathbold{x}$ is a Gaussian codeword of length $mn$. This implies that, for a non Gaussian $p_\mathbold{x}$, an attack strategy $p_\mathbold{v|z}^o\notin\mathcal{C}$ may exist, that achieves 
\begin{equation}
    f(p_\mathbold{x}, p_\mathbold{v|z}^o) \leq f(p_\mathbold{x}, p_\mathbold{v|z}^\star)\,.  
    \label{eqn:chain2}
\end{equation}
Therefore, from (\ref{eqn:chain1})--(\ref{eqn:chain2}) we can derive the following relation
\begin{multline}
    f(p_\mathbold{x}, p_\mathbold{v|z}^o) \leq f(p_\mathbold{x}, p_\mathbold{v|z}^\star) \\ = f(p_\mathbold{x}^\star, p_\mathbold{v|z}^\star) \leq f(p_\mathbold{x}^\star, p_\mathbold{v|z}^o)\,.
    \label{eqn:chainf}
\end{multline}

When the signal $\mathbold{x}$ has symbols from a finite set, the transmission strategy differs from $p_\mathbold{x}^\star$, and the optimal attack strategy distribution $p_\mathbold{v|z}^o$ is not known. 
Hence, the receiver only knows the statistics of the authentic signal, and cannot make assumptions on the attack strategy chosen by $\rm{E}$ nor has information on the channels $\rm{A}\rightarrow \rm{E}$ and $\rm{E}\rightarrow \rm{B}$. In this case, the \ac{lrt} detection method no longer applies, and a possible solution is to use the \ac{glrt} \cite{signalProcKay,GLRTopt?}, i.e., the detection problem is given by
\begin{equation}
    G' = (\mathbold{r}-\mathbold{Ax})^\mathrm{H}\mathbold{K_\mathrm{B}}^{-1}(\mathbold{r}-\mathbold{Ax}) \underset{H_0}{\overset{H_1}{\gtrless}} \theta\,.
\end{equation}

\section{Numerical Results}
\label{par:results}

In this Section we will illustrate the performance obtained for both \ac{lrt} and \ac{glrt}, when either Gaussian or finite-cardinality signaling are used, under the following scenario:
\begin{itemize}
    \item channel matrices $\mathbold{A}$ and $\mathbold{F}$ take into account only the delays, that is the propagation times of each signal $\bar{\mathbold{x}}_i$ from the $i$-th \ac{sv} to Eve and Bob, therefore we neglect other possible channel phenomena;
    \item the attack strategy $p_\mathbold{v|z}^\star$ is that described in Section \ref{par:min}.
\end{itemize}
Moreover, in a typical \ac{gnss} spoofing scenario, the power of the spoofing signal $\mathbold{v_0}$ at the receiver is typically larger than that of the legitimate signal; thus, the \ac{snr} on the attack signal will be significantly larger than with the authentic one. Assuming that an automatic gain control (AGC) at Bob's front-end can scale the received signal to nearly constant amplitude, the receiver noise variance will be correspondingly reduced, in the attack case, allowing the spoofer some margin to shape $\mathbold{K_{\eta}} = \mathbold{K_\mathrm{B}}$.
Similarly to \eqref{eqn:lambda0}, we define the received spoofer \ac{snr} as
\begin{equation}
    \Lambda_\mathrm{AE} = \frac{M_x}{\sigma^2_\mathrm{E}} \, \frac{\|\mathbold{F}\|^2_F}{mn}\,,   
\end{equation}

\begin{figure}
\centering
\includegraphics[width=\columnwidth]{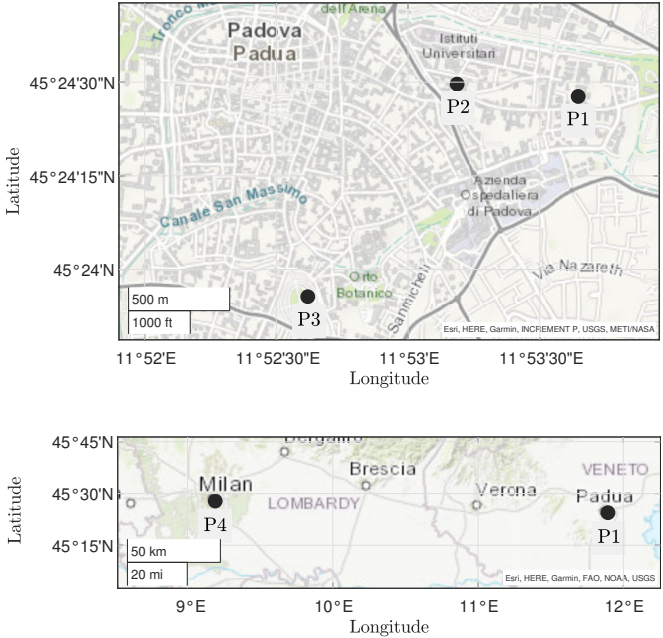}
\caption[]{Map of the considered positions. Three are located in Padua, Italy: DEI (P1), Paolotti street (P2), and square of Prato della Valle (P3). One located in Milan, Italy: Duomo square (P4).}
\label{fig:mappa}
\end{figure}

\begin{table}
    \centering
    \caption{Considered positions and their coordinates}
    \begin{tabularx}{\columnwidth}{l|l|l}
    \toprule
         & LLH coordinates  & Landmark \\
    \midrule
       P1 & $45.4077^\circ$ N, $11.8941^\circ$ E, 12 m & DEI, Padua, Italy \\
       P2 & $45.4079^\circ$ N, $11.8860^\circ$ E, 12 m & via Paolotti, Padua, Italy\\
       P3 & $45.3980^\circ$ N, $11.8766^\circ$ E, 12 m & Prato della Valle, Padua, Italy \\
       P4 & $45.4641^\circ$ N, $9.1903^\circ$ E, 120 m & Duomo square, Milan, Italy \\
        \bottomrule
    \end{tabularx} \label{tab:landmarks}
\end{table}

\begin{table}[t]
    \centering
    \caption{Distances between Eve and target position}
    \begin{tabular}{l|l|l}
    \toprule
        Eve position & Target position  & Distance \\
    \midrule
       P1 & P2 & 607 \,m \\
       P1 & P3 & 1.7\;\,  Km \\
       P1 & P4 & 211 Km\\
        \bottomrule
    \end{tabular} \label{tab:distances}
\end{table}
In the analyzed scenarios, we will take into account the delay vector $\mathbold{\tau_\mathrm{f}}$ associated with the position of our department (DEI) (P1), located in Padua, Italy, and the delay vector $\mathbold{\tau_\mathrm{E}}$ associated to three different positions, two of which located in Padua, Italy, namely a house in Paolotti street (P2), and the square of Prato della Valle (P3), and the Duomo square (P4), located in Milan, Italy. 
Figure \ref{fig:mappa} illustrates the four positions. Delays' measurements have been collected the 14th December 2022, at 12.00 AM. Table \ref{tab:landmarks} outlines each considered position with their coordinates, while Table \ref{tab:distances} summarizes the distances between target and Eve's positions.

\subsection{Gaussian Signaling}

In this Section performance is evaluated when the transmitted signal $\mathbold{x}$ is a Gaussian codeword of length $n$, considering a \ac{lrt} detection method.

\begin{figure}
\centering
\includegraphics[width=1.06\columnwidth]{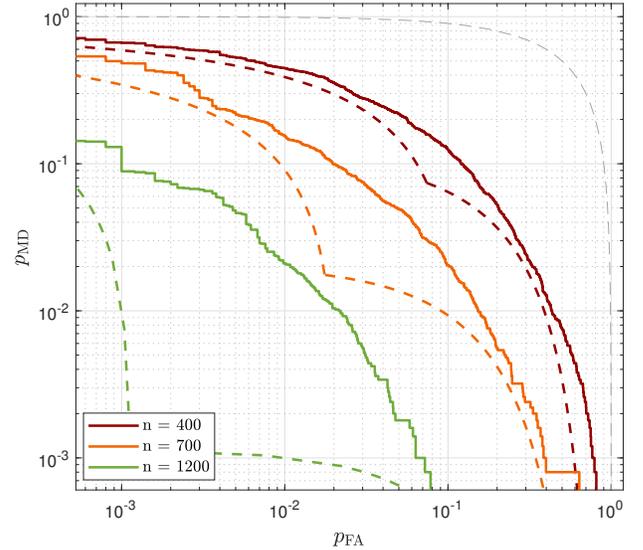}
\caption[]{\ac{det} curves for the \ac{lrt} detection method (solid lines) and \ac{kl} divergence bounds (dashed lines) for different values of $n$, when $p_\mathbold{x}$ is Gaussian distributed, $M_x=1$, $m=5$, $\mathbold{\tau_\mathrm{f}}$ associated to P1, $\mathbold{\tau_\mathrm{E}}$ associated to P2, $\Lambda_\mathrm{AB}=-25$ dB, and $\Lambda_\mathrm{AE}=-10$ dB. The gray dashed line represents the trivial limit case in which the decision is taken tossing a biased coin.}
\label{fig:LRTgaussn}
\end{figure}
Fig. \ref{fig:LRTgaussn} shows the \ac{det} curves for the \ac{lrt} detection method (solid lines) and the performance bounds (dashed lines) derived from the \ac{kl} divergence for different values of $n$, considering the signals coming from $m=5$
\acp{sv}, $M_x=1$, $\Lambda_\mathrm{AB}=-25$ dB, $\Lambda_\mathrm{AE}=-10$ dB, when $\mathbold{\tau_\mathrm{E}}$ and $\mathbold{\tau_\mathrm{f}}$ collect the delays associated to P1 and P2, respectively. The bound follows the same trend of (\ref{eqn:div3terms}), thus it is representative of the actual performance. The bound is tight for $n=400$, while the gap between it and the actual performance grows as the value of $n$ rises. Indeed, as $n$ decreases, the \ac{det} curves move quickly towards the gray dashed line, which represents the trivial limit case, in which the decision is taken without looking at the signal but tossing a biased coin.
Thus, the obtained value of $p_\mathrm{MD}$ for given values of $p_\mathrm{FA}$ can be reduced by increasing the value of $n$, as Fig.~\ref{fig:LRTgaussn} shows. When $n=1200$, the value of $p_\mathrm{MD}$ decreases by approximately one order of magnitude with respect to the case with $n=400$.
Moreover, we note that the bounds given by the \ac{kl} divergence are symmetric, as proven in Appendix \ref{app:simmKL}.


\begin{figure}
\centering
\subfloat[]{\includegraphics[width=1.06\columnwidth]{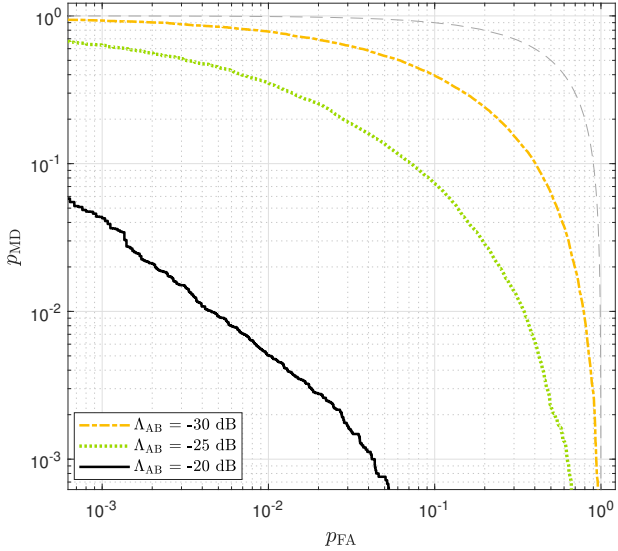}}
    \label{fig:LRTgausssigmaB}
\subfloat[]{\includegraphics[width=1.06\columnwidth]{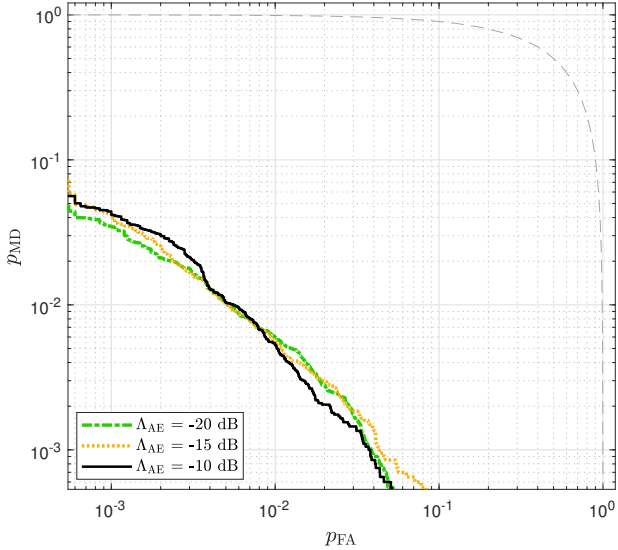}}
    \label{fig:LRTgausssigmaE}
    \vspace{0.2cm}
\caption[]{\ac{det} curves for the \ac{lrt} detection method for different values of (a) $\Lambda_\mathrm{AB}$ and (b) $\Lambda_\mathrm{AE}$, when $p_\mathbold{x}$ is Gaussian distributed, $M_x=1$, $m=5$, $n=500$, $\mathbold{\tau_\mathrm{f}}$ associated to P1, and $\mathbold{\tau_\mathrm{E}}$ associated to P2. In (a) $\Lambda_\mathrm{AE}=-10$ dB, and (b) $\Lambda_\mathrm{AB}=-20$ dB.}
\label{fig:LRTgausssigmaBE}
\end{figure}
Fig. \ref{fig:LRTgausssigmaBE} shows the \ac{det} curves of the \ac{lrt} detection method for different values of (a) $\Lambda_\mathrm{AB}$ and (b) $\Lambda_\mathrm{AE}$, considering the signals coming from $m=5$
\acp{sv}, $M_x=1$, in (a) $\Lambda_\mathrm{AE}=-10$ dB, in (b) $\Lambda_\mathrm{AB}=-20$ dB,
when $\mathbold{\tau_\mathrm{E}}$ and $\mathbold{\tau_\mathrm{f}}$ collect the delays associated to P1 and P2, respectively. It can be seen that the curves follow the behavior of (\ref{eqn:div3terms}); in fact, the curves rise when $\Lambda_\mathrm{AB}$ decreases, while remaining unchanged for different values of $\Lambda_\mathrm{AE}$. As a result, the performance of the \ac{lrt} verification mechanism is independent of the attacker's \ac{snr}, as long as $\Lambda_\mathrm{AE} > \Lambda_\mathrm{AB}$.

\begin{figure}
\centering
\includegraphics[width=1.06\columnwidth]{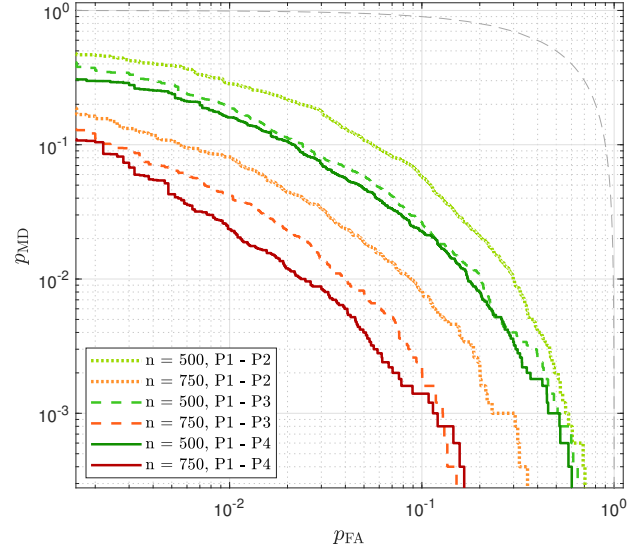}
\caption[]{\ac{det} curves for the \ac{lrt} detection method for different values of $n$ and position pairs, when $p_\mathbold{x}$ is Gaussian distributed, $M_x=1$, $m=5$, $\Lambda_\mathrm{AB}=-25$ dB, and $\Lambda_\mathrm{AE}=-10$ dB.}
\vspace{-0.1cm}
\label{fig:LRTgausspos}
\end{figure}

Fig. \ref{fig:LRTgausspos} shows the \ac{det} curves for the \ac{lrt} detection method for different values of $n$ and position pairs, illustrated in Fig. \ref{fig:mappa}, when $M_x=1$, $m=5$, $\Lambda_\mathrm{AB}=-25$ dB, and $\Lambda_\mathrm{AE}=-10$ dB. In each tested scenario, the delays vector $\mathbold{\tau_\mathrm{f}}$ is associated to P1, while the vector $\mathbold{\tau_\mathrm{E}}$ is associated with three different positions, i.e., (from the closest to P1 to furthest): P2, P3, and P4. We can see that the curves move away from perfect distinguishability (the lower left corner) when the distance between the position associated with $\mathbold{\tau_\mathrm{E}}$ and $\mathbold{\tau_\mathrm{f}}$ decreases. Therefore, the attack performance degrades as the attack's target position moves farther away from the attacker's actual position. Finally, we note that, as for Fig. \ref{fig:LRTgaussn}, the defense performance improves (i.e., the curves move towards the bottom left corner) as $n$ increases.

\subsection{Practical Modulation Schemes}

As discussed in Section \ref{par:GaussFin}, switching from Gaussian signaling to finite-cardinality signaling does not affect the performance of the verification mechanism, when the attack strategy belongs to $\mathcal{C}$. This behavior will be demonstrated in this paragraph through simulation results. In the following, a \ac{bpsk} modulation will be considered (i.e., having cardinality $M = 2$).

\begin{figure}
\centering
\includegraphics[width=1.06\columnwidth]{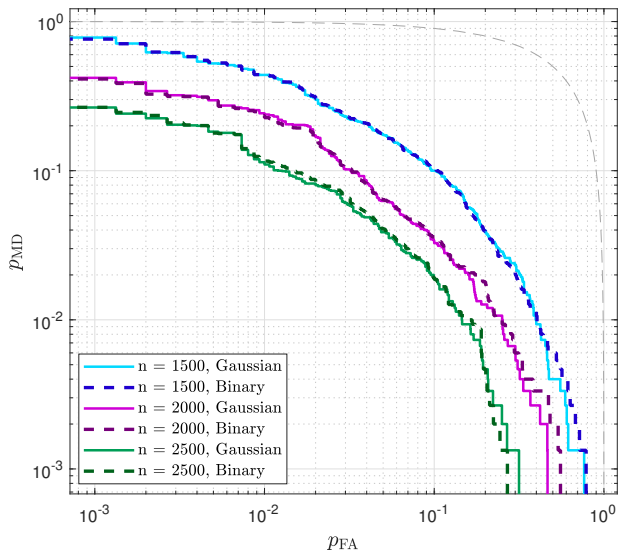}
\caption[]{\ac{det} curves for the \ac{glrt} detection method for different values of $n$, when $p_\mathbold{x}$ is Gaussian distributed (solid lines) and when $p_\mathbold{x}$ is a Binary distribution (dashed lines), $M_x=1$, $m=9$, $\mathbold{\tau_\mathrm{f}}$ associated to P1, $\mathbold{\tau_\mathrm{E}}$ associated to P4, $\Lambda_\mathrm{AB}=-20$ dB, and $\Lambda_\mathrm{AE}=-10$ dB.}
\label{fig:GLRTbingauss}
\end{figure}

Fig. \ref{fig:GLRTbingauss} shows the \ac{det} curves for the \ac{glrt} detection method for different values of $n$, when $p_\mathbold{x}$ is Gaussian distributed (solid lines) and $p_\mathbold{x}$ is a Binary distribution (dashed lines), $M_x=1$, $m=9$, $\mathbold{\tau_\mathrm{f}}$ and $\mathbold{\tau_\mathrm{E}}$ collect the delays associated to P1 and P4, respectively. These results have been obtained for $\Lambda_\mathrm{AB}=-20$ dB and $\Lambda_\mathrm{AE}=-10$ dB, so we are in the low-\ac{snr} regime. As can be seen, the curves obtained with Gaussian and binary signaling are very close, confirming the results of Section \ref{par:GaussFin}.
 Moreover, it is seen that \ac{glrt} is effective in detecting spoofing, even without any prior knowledge of the spoofer strategy. For practically meaningful values for $p_\mathrm{FA}$ and $p_\mathrm{MD}$, we should consider longer signals, so a higher value of $n$ with respect to the \ac{lrt} case. 
 Clearly, by observing more samples before taking a decision, the decision will be more accurate, but this requires to buffer and process more data. Furthermore, this leads to a longer \ac{tta}. Thus, the observation period must be chosen as a trade-off between the computational resources of the device, the desired \ac{tta}, and the desired performance in terms of \ac{det}. 

\begin{figure}
\centering
\includegraphics[width=1.06\columnwidth]{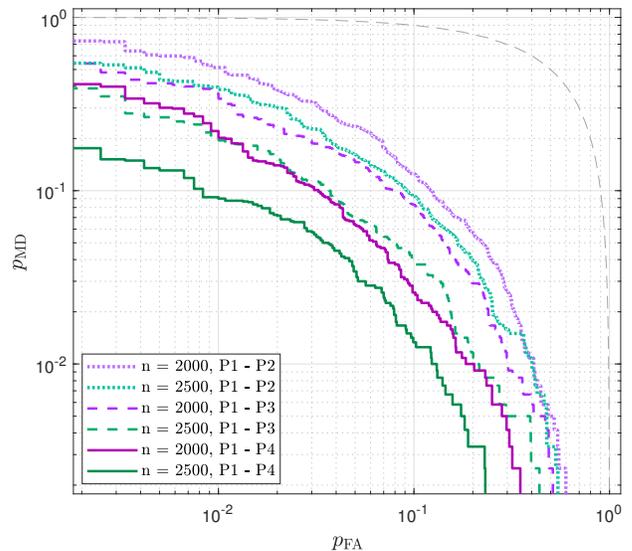}
\caption[]{\ac{det} curves for the \ac{glrt} detection method for different values of $n$ and position pairs, when $p_\mathbold{x}$ is Gaussian distributed, $M_x=1$, $m=9$, $\Lambda_\mathrm{AB}=-20$ dB, and $\Lambda_\mathrm{AE}=-10$ dB.}
\label{fig:GLRTgausspos}
\end{figure}
Fig. \ref{fig:GLRTgausspos} shows the \ac{det} curves for the \ac{glrt} detection method for different values of $n$ and position pairs, when $p_\mathbold{x}$ is Gaussian distributed, $M_x=1$, $m=9$, $\Lambda_\mathrm{AB}=-20$ dB, and $\Lambda_\mathrm{AE}=-10$ dB. In each tested scenario, the delay vector $\mathbold{\tau_\mathrm{f}}$ is associated to P1, while vector $\mathbold{\tau_\mathrm{E}}$ is associated with P2, P3, and P4. As for the \ac{lrt} case, the attack performance improves when the distance between the position associated with $\mathbold{\tau_\mathrm{E}}$ and $\mathbold{\tau_\mathrm{f}}$ decreases. Therefore, also for the \ac{glrt}, the missed detection probability decreases, for a fixed false alarm probability, as the attack's target position moves farther away from the attacker's actual position.

\section{Conclusion}
\label{par:conclusion}
In this paper we have proposed a general model to characterize the spoofing detection problem in \acp{gnss} when the spoofer can observe the legitimate signal, abstracting from the specific modulation formats and cryptographic mechanisms. We have shown that effective detection can be achieved by relying only on the combination of signals from multiple \acp{sv} and on the diversity between the attacker position and the intended forged position. We have also investigated a class of attack strategies based on the statistics of the transmitted and received signals, which are shown to be optimal to minimize the \ac{kl} divergence metric. The optimal attack strategy turned out to be a proper linear transformation of the signal received at the attacker position, combined with an appropriately tuned independent additive white Gaussian noise.
We have derived a lower bound on the \ac{kl} divergence, which depends only on the total length of the transmitted signals, on the \ac{snr} of the legitimate channel, and on the difference between the forged and the legitimate channel matrices. Moreover, we have discussed the results obtained in relation to different modulation schemes; we have shown that when the attack strategy is the optimal one, with Gaussian or finite-cardinality signaling we get the same performance in terms of \ac{kl} divergence. 
Then, we found a Nash equilibrium of the attack-defense scheme  deriving the optimal defense strategy against the above-mentioned attack, which, in turn, is described by a Gaussian distribution. 
Finally, the performance of the detection schemes against the proposed attack has been analyzed through numerical simulations considering two verification mechanisms: \ac{lrt} and \ac{glrt}.


\appendices
\section{Symmetry of the K-L divergence}
\label{app:simmKL}
In this Appendix we provide a proof of the symmetry of the \ac{kl} divergence when the attacker uses the optimal attack strategy $p_\mathbold{v|z}^\star$, presented in Section \ref{par:min}, and $\mathbold{K_{\eta}} = \mathbold{K_\mathrm{B}}$, that is
\begin{equation}
    \mathbb{D}(p_{\mathbold{xv}^\star}\|p_{\mathbold{xy}}) =  \mathbb{D}(p_{\mathbold{xy}}\| p_{\mathbold{xv}^\star} )\,.
    \label{eqn:symDiv}
\end{equation}

For any $p_\mathbold{v|z}\in\mathcal{C}$, $\mathbb{D}(p_{\mathbold{xy}}\| p_{\mathbold{xv}})$ can be computed following the same procedure as in (\ref{eqn:divkx}), therefore we get
\begin{equation}
\begin{split}
    \mathbb{D}&(p_{\mathbold{xy}}\| p_{\mathbold{xv}}) = \frac{1}{2}\bigg[\log\frac{|\mathbold{K_\eta}|}{|\mathbold{K_\mathrm{B}}|}+\tr(\mathbold{K_\mathrm{B}K_\eta}^{-1}) \\ &+\tr\bigg((\mathbold{A-B})\mathbold{K_x}(\mathbold{A-B})^\mathrm{H}\mathbold{K_\mathrm{B}}^{-1} \bigg) - (n+\delta_\mathrm{f}) \bigg]\\
=&\, \frac{1}{2} \left[ \sum_{i=1}^{n+\delta_\mathrm{f}}\left(\frac{\sigma^2_\mathrm{B}}{\lambda_i}-\log\left(\frac{\sigma^2_\mathrm{B}}{\lambda_i} \right)\right) -(n+\delta_\mathrm{f}) \right] \\ &+ \frac{1}{2\sigma^2_\mathrm{B}}\tr\left((\mathbold{A-B})\mathbold{K_x}(\mathbold{A-B})^\mathrm{H}\right)
\end{split}
\end{equation}
When the attack strategy is $p_\mathbold{v|z}^\star$ and $\mathbold{K_{\eta}} = \mathbold{K_\mathrm{B}}$, we have that $\lambda_i = \sigma^2_\mathrm{B}, \forall i \in \{1, 2, ...,  n+\delta_\mathrm{f}\}$. Therefore, we obtain
\begin{multline}
    \mathbb{D}(p_{\mathbold{xy}} \| p_{\mathbold{xv^\star}}) = \frac{1}{2\sigma^2_\mathrm{B}}\tr\left((\mathbold{A-B^\star})\mathbold{K_x}(\mathbold{A-B^\star})^\mathrm{H}\right)
    \\ = \mathbb{D}_{min}(\mathbold{B}^\star )
    = \mathbb{D}(p_{\mathbold{xv^\star}} \| p_{\mathbold{xy}})\,.
    \end{multline}

\ifCLASSOPTIONcaptionsoff
  \newpage
\fi
\balance

\bibliography{biblioGenMod.bib}{}
\bibliographystyle{IEEEtran}

\end{document}